\UseRawInputEncoding
\documentclass[pra,twocolumn,shownopacs,amssymb]{revtex4}
\usepackage{graphicx,psfrag,amsmath,amssymb}
    
\begin{document}

\title{Three-body problem in a multiband Hubbard model}

\author{M. Iskin}
\affiliation{Department of Physics, Ko\c{c} University, Rumelifeneri Yolu, 
34450 Sar\i yer, Istanbul, Turkey}

\date{\today}

\begin{abstract}

We consider the three-body problem in a generic multiband lattice, and analyze the 
dispersion of the trimer states that are made of two spin-$\uparrow$ fermions and 
a spin-$\downarrow$ fermion due to an onsite attraction in between. 
Based on a variational approach, we first obtain the exact solution in the form of a 
set of coupled integral equations, and then reduce it to an eigenvalue problem. 
As an illustration we apply our theory to the sawtooth lattice, and numerically show 
that energetically-stable trimers are allowed in a two-band setting, which is in sharp 
contrast with the single-band linear-chain model. In particular we also reveal that 
the trimers have a nearly-flat dispersion when formed in a flat band, which is unlike 
the highly-dispersive spectrum of its dimers.

\end{abstract}

\maketitle

\section{Introduction}
\label{sec:intro}

The Hubbard model and its numerous extensions are major playgrounds for studying 
central research problems in solid-state, condensed-matter, and atomic and molecular 
physics, particularly when the role played by the interactions is 
indispensable~\cite{tasaki, arovas21, Qin21}. 
Despite their drastic simplifications, these models have been successfully used to 
elucidate and predict complex phenomena ranging from quantum magnetism, 
superconductivity and superfluidity to metal-insulator transition, charge-density waves, 
superfluid-Mott insulator transition and supersolidity. There is no doubt that the 
significance of Hubbard-type models to quantum many-body physics is akin to 
that of the Ising model to statistical mechanics or the fruit fly to molecular 
biology~\cite{arovas21, Qin21}. 

Nowadays these models are routinely used to characterize the ultracold-atom based 
quantum simulators that are constructed by trapping a gas of atoms (that obey Fermi 
or Bose statistics or a mixture of both) on optical lattice potentials~\cite{esslinger10, gross17}. 
By designing tailor-cut experiments that mimic Hubbard-type simplistic models, 
the ultimate hope in this field is to gain deeper understanding on specific problems 
that are theoretically and sometimes numerically intractable. In contrast with the 
many-body problems where much of the phase diagrams remain controversial, 
exactly-solvable few-body problems stand out as ideal testbeds for new theoretical
ideas and approaches. For instance the creation of long-sought Efimov trimers 
with three identical bosons in continuum, i.e., without the lattice, is one of the major 
breakthroughs in modern atomic
physics~\cite{kraemer06, zaccanti09, pollack09, braaten06, greene17, naidon17, zinner14}, 
which stimulated tons of trimer research with fermions as well, 
e.g., see~\cite{greene17, naidon17, km07, shi14, cui14, ji20}.

Motivated by the recent creation of Kagome~\cite{jo12, nakata12, li18} and 
Lieb~\cite{diebel16, kajiwara16, ozawa17} lattices, and ongoing activity 
in strongly-correlated electrons or atoms in a flat 
band~\cite{tasaki98, parameswaran13, liu14, leykam18, balents20}, here 
we consider the three-body problem in a generic multiband Hubbard model, 
and discuss the dispersion of the trimer states that are made of two 
spin-$\uparrow$ fermions and a spin-$\downarrow$ fermion. This is achieved through 
a variational approach and by reducing its exact solutions to an eigenvalue problem. 
As an illustration we apply our theory to the sawtooth lattice with a two-point basis,
and show that the trimer states are allowed in a broad range of model parameters.
This finding is in sharp contrast with the single-band linear-chain model and it is in 
very good agreement with the recent DMRG results~\cite{orso21}. 
In addition we find that the trimers have a nearly-flat dispersion with a negligible 
bandwidth when formed in a flat band. This is quite peculiar given the highly-dispersive 
spectrum of the two-body bound states (dimers) in the same system. 

The rest of the text is organized as follows. In Sec.~\ref{sec:va} we first introduce the 
model Hamiltonian and the variational ansatz for the three-body problem, and then
derive a set of coupled integral equations. In Sec.~\ref{sec:num} we recast the integral
equations as an eigenvalue problem for the dispersion of the bound states. 
In Sec.~\ref{sec:saw} we apply our theory to the sawtooth lattice and discuss the
binding energy of its trimer states in a broad range of model parameters. 
In Sec.~\ref{sec:conc} we end the paper with a brief summary of our conclusions.

\section{Variational Approach}
\label{sec:va}

The Hubbard model is one of the simplest descriptions of interacting fermions in a lattice
with only two terms 
$
H = \sum_\sigma H_\sigma + H_{\uparrow\downarrow}
$
contributing to its Hamiltonian. The first term
$
H_\sigma = -\sum_{Si; S'i'} t_{Si; S'i'}^\sigma c_{S i \sigma}^\dagger c_{S' i' \sigma} 
$
describes the kinetic energy of spin-$\sigma$ fermions, where the operator $c_{S i \sigma}^\dagger$ 
creates a spin-$\sigma$ fermion in the unit cell $i$ at the sublattice $S$, and the hopping
parameter $t_{Si; S'i'}^\sigma$ corresponds to the transfer energy that is gained/lost by 
the particle when it hops from site $S' i'$ to site $S i$. The second term 
$
H_{\uparrow\downarrow} = - U \sum_{S i} \rho_{S i \uparrow} \rho_{S i \downarrow}
$
describes the potential energy, i.e., onsite attraction, between spin-$\uparrow$ and 
spin-$\downarrow$ particles, where the operator
$
\rho_{S i \sigma} = c_{S i \sigma}^\dagger c_{S i \sigma}
$
counts the number of spin-$\sigma$ fermions at site $S i$ and the interaction parameter 
$U \ge 0$ measures the strength of the attraction.

In order to take advantage of the discrete-translational symmetry of the lattice, it is convenient 
to express the Hamiltonian in the first Brillouin zone (BZ) through the Fourier expansion 
$
c_{S i \sigma}^\dagger = \frac{1}{\sqrt{N_c}} \sum_\mathbf{k} 
e^{-i \mathbf{k} \cdot \mathbf{r_{S i}}} c_{S \mathbf{k} \sigma}^\dagger.
$
Here the integer $N_c$ is the number of unit cells in the lattice, the wave vector 
$\mathbf{k} \in \mathrm{BZ}$ is the crystal momentum (in units of $\hbar = 1$), the vector 
$\mathbf{r_{S i}}$ is the position of the site $S i$, and the operator 
$c_{S \mathbf{k} \sigma}^\dagger$ creates a spin-$\sigma$ fermion in the sublattice $S$ with 
momentum $\mathbf{k}$. The total number of lattice sites is given by $N = N_b N_c$ where
$N_b$ is the number of basis sites (sublattices) in a unit cell. Since the resultant 
$N_b \times N_b$ Bloch matrix is diagonal in the band representation 
(for a given $\mathbf{k}$), the spin-$\sigma$ Hamiltonian can be expressed as 
\begin{align}
\label{eqn:Hsigma}
H_\sigma = \sum_{n \mathbf{k}} \varepsilon_{n\mathbf{k}\sigma}
c_{n \mathbf{k} \sigma}^\dagger c_{n \mathbf{k} \sigma},
\end{align}
where the operator $c_{n \mathbf{k} \sigma}^\dagger$ creates a spin-$\sigma$ fermion 
in the Bloch band $n$ with momentum $\mathbf{k}$ and energy $\varepsilon_{n\mathbf{k}\sigma}$.
We denote the corresponding Bloch state as
$
| n \mathbf{k} \sigma \rangle = c_{n \mathbf{k} \sigma}^\dagger | 0 \rangle,
$
whose sublattice projections $n_{S \mathbf{k} \sigma} = \langle S| n \mathbf{k} \sigma \rangle$ 
links the operators in different basis, i.e.,
$
c_{n \mathbf{k} \sigma}^\dagger = \sum_S n_{S \mathbf{k} \sigma} 
c_{S \mathbf{k} \sigma}^\dagger.
$
Here the state $| 0 \rangle$ corresponds to the vacuum of particles.
Similarly a compact way to express the interaction Hamiltonian is~\cite{iskin21}
\begin{align}
\label{eqn:Hupdown}
H_{\uparrow\downarrow} = \frac{1}{N_c} 
\sum_{\substack{nmn'm' \\ \mathbf{k}\mathbf{k'}\mathbf{q}}}
V_{n'm'\mathbf{k'}}^{nm\mathbf{k}}(\mathbf{q})
b_{nm}^\dagger(\mathbf{k}, \mathbf{q})
b_{n'm'}(\mathbf{k'}, \mathbf{q}),
\end{align}
where the operator
$
b_{nm}^\dagger (\mathbf{k}, \mathbf{q}) = c_{n,\mathbf{k}+\frac{\mathbf{q}}{2}, \uparrow}^\dagger
c_{m,-\mathbf{k}+\frac{\mathbf{q}}{2}, \downarrow}^\dagger
$
creates a pair of fermions with relative momentum $2\mathbf{k}$ and total 
momentum $\mathbf{q}$, and 
$
V_{n'm'\mathbf{k'}}^{nm\mathbf{k}}(\mathbf{q}) = - U\sum_S 
n_{S, \mathbf{k}+\frac{\mathbf{q}}{2}, \uparrow}^*
m_{S, -\mathbf{k}+\frac{\mathbf{q}}{2}, \downarrow}^*
{m'}_{S, -\mathbf{k'}+\frac{\mathbf{q}}{2}, \downarrow}
{n'}_{S, \mathbf{k'}+\frac{\mathbf{q}}{2}, \uparrow}
$
characterizes the long-range interactions in momentum space.

In this paper we solve the Schr\"odinger equation 
$
H | \Psi_\mathbf{q} \rangle = E_{3b}^\mathbf{q} | \Psi_\mathbf{q} \rangle,
$
and obtain the exact solutions to the three-body problem through a variational 
approach that is based on the following ansatz
\begin{align}
\label{eqn:Psiq}
| \Psi_\mathbf{q} \rangle = \sum_{nm\ell \mathbf{k_1} \mathbf{k_2}} 
\alpha_{nm\ell}^{\mathbf{k_1} \mathbf{k_2}} (\mathbf{q})
c_{n \mathbf{k_1} \uparrow}^\dagger
c_{m \mathbf{k_2} \uparrow}^\dagger 
c_{\ell, \mathbf{q}-\mathbf{k_1}-\mathbf{k_2}, \downarrow}^\dagger
| 0 \rangle.
\end{align}
This ansatz represents the three-body bound states for a given total momentum
$\mathbf{q}$ of the particles, and its complex variational parameters 
$
\alpha_{nm\ell}^{\mathbf{k_1} \mathbf{k_2}} (\mathbf{q})
$
are determined through the functional minimization of 
$
\langle \Psi_\mathbf{q} | H - E_{3b}^\mathbf{q} | \Psi_\mathbf{q} \rangle.
$
The $\mathbf{q}$-dependence of
$
\alpha_{nm\ell}^{\mathbf{k_1} \mathbf{k_2}} (\mathbf{q})
$
is suppressed in some parts of the text for the simplicity of the presentation.
For instance the normalization condition is
$
\langle \Psi_\mathbf{q} | \Psi_\mathbf{q} \rangle = 
\sum_{nm\ell \mathbf{k_1} \mathbf{k_2}} 
\big[
|\alpha_{nm\ell}^{\mathbf{k_1} \mathbf{k_2}}|^2 
- (\alpha_{mn\ell}^{\mathbf{k_2} \mathbf{k_1}})^* 
\alpha_{nm\ell}^{\mathbf{k_1} \mathbf{k_2}}
\big].
$
By plugging Eq.~(\ref{eqn:Psiq}) into the Schr\"odinger equation that is governed by the
Hamiltonians given in Eqs.~(\ref{eqn:Hsigma}) and~(\ref{eqn:Hupdown}), we find
\begin{align*}
\langle H_\uparrow \rangle &= 
\sum_{\substack{nm\ell \\ \mathbf{k_1} \mathbf{k_2}}} 
\big[
|\alpha_{nm\ell}^{\mathbf{k_1} \mathbf{k_2}}|^2 
- (\alpha_{mn\ell}^{\mathbf{k_2} \mathbf{k_1}})^* 
\alpha_{nm\ell}^{\mathbf{k_1} \mathbf{k_2}}
\big] 
(\varepsilon_{n \mathbf{k_1} \uparrow} + \varepsilon_{m \mathbf{k_2} \uparrow}),
\\
\langle H_\downarrow \rangle &= 
\sum_{\substack{nm\ell \\ \mathbf{k_1} \mathbf{k_2}}} 
\big[
|\alpha_{nm\ell}^{\mathbf{k_1} \mathbf{k_2}}|^2 
- (\alpha_{mn\ell}^{\mathbf{k_2} \mathbf{k_1}})^* 
\alpha_{nm\ell}^{\mathbf{k_1} \mathbf{k_2}}
\big] 
\varepsilon_{\ell, \mathbf{q} - \mathbf{k_1} - \mathbf{k_2}, \downarrow},
\\
\langle H_{\uparrow\downarrow} \rangle &= 
-\frac{U}{N_c} \sum_{\substack{nm\ell n'm' \\ S \mathbf{k_1} \mathbf{k_2} \mathbf{k_3}}}
\big[ \nonumber \\
&\quad\,\quad
(\alpha_{n'mm'}^{\mathbf{k_3} \mathbf{k_2}})^* \alpha_{nm\ell}^{\mathbf{k_1} \mathbf{k_2}} 
{n'}_{S \mathbf{k_3} \uparrow}^*
{m'}_{S \mathbf{Q_{32}} \downarrow}^*
\ell_{S \mathbf{Q_{12}}  \downarrow}
n_{S \mathbf{k_1} \uparrow} \nonumber \\
&\quad -
(\alpha_{mm'n'}^{\mathbf{k_2} \mathbf{k_3}})^* \alpha_{nm\ell}^{\mathbf{k_1} \mathbf{k_2}}
{m'}_{S \mathbf{k_3} \uparrow}^*
{n'}_{S \mathbf{Q_{23}} \downarrow}^*
\ell_{S \mathbf{Q_{12}}  \downarrow}
n_{S \mathbf{k_1} \uparrow} \nonumber \\
&\quad -
(\alpha_{n'mm'}^{\mathbf{k_3} \mathbf{k_1}})^* \alpha_{nm\ell}^{\mathbf{k_1} \mathbf{k_2}} 
{n'}_{S \mathbf{k_3} \uparrow}^*
{m'}_{S \mathbf{Q_{31}} \downarrow}^*
\ell_{S \mathbf{Q_{12}}  \downarrow}
m_{S \mathbf{k_2} \uparrow} \nonumber \\
&\quad + 
(\alpha_{nm'n'}^{\mathbf{k_1} \mathbf{k_3}})^* \alpha_{nm\ell}^{\mathbf{k_1} \mathbf{k_2}} 
{m'}_{S, \mathbf{k_3}, \uparrow}^*
{n'}_{S \mathbf{Q_{13}} \downarrow}^*
\ell_{S \mathbf{Q_{12}}  \downarrow}
m_{S \mathbf{k_2} \uparrow} 
\big],
\end{align*}
where we define
$
\mathbf{Q_{ij}} = \mathbf{q} - \mathbf{k_i} - \mathbf{k_j}
$
as a shorthand notation.
Thus, by setting 
$
\partial \langle H - E_{3b}^\mathbf{q} \rangle / \partial (\alpha_{nm\ell}^{\mathbf{k_1} \mathbf{k_2}})^* = 0
$ 
for a given $\mathbf{q}$, we obtain a set of coupled integral equations that must be satisfied 
by $E_{3b}^\mathbf{q}$ and $\alpha_{nm\ell}^{\mathbf{k_1} \mathbf{k_2}}(\mathbf{q})$ 
simultaneously, i.e.,
\begin{align}
\big( \varepsilon_{n \mathbf{k_1} \uparrow} + \varepsilon_{m \mathbf{k_2} \uparrow} 
&+ \varepsilon_{\ell \mathbf{Q_{12}} \downarrow} - E_{3b}^\mathbf{q} \big)
\big(\alpha_{nm\ell}^{\mathbf{k_1} \mathbf{k_2}} - \alpha_{mn\ell}^{\mathbf{k_2} \mathbf{k_1}}\big)
\nonumber \\
= \frac{U}{N_c} \sum_{n'm'S\mathbf{k_3}} 
\big(
&\alpha_{n'mm'}^{\mathbf{k_3} \mathbf{k_2}} 
n_{S \mathbf{k_1} \uparrow}^* \ell_{S \mathbf{Q_{12}} \downarrow}^*
{m'}_{S \mathbf{Q_{32}}  \downarrow} {n'}_{S \mathbf{k} \uparrow}
\nonumber \\
-&
\alpha_{n'nm'}^{\mathbf{k_3} \mathbf{k_1}} 
m_{S \mathbf{k_1} \uparrow}^* \ell_{S \mathbf{Q_{12}} \downarrow}^*
{m'}_{S \mathbf{Q_{31}}  \downarrow} {n'}_{S \mathbf{k} \uparrow}
\nonumber \\
-&
\alpha_{mn'm'}^{\mathbf{k_2} \mathbf{k_3}} 
n_{S \mathbf{k_1} \uparrow}^* \ell_{S \mathbf{Q_{12}} \downarrow}^*
{m'}_{S \mathbf{Q_{23}} \downarrow} {n'}_{S \mathbf{k} \uparrow}
\nonumber \\
+ &
\alpha_{nn'm'}^{\mathbf{k_1} \mathbf{k_3}} 
m_{S \mathbf{k_2} \uparrow}^* \ell_{S \mathbf{Q_{12}} \downarrow}^*
{m'}_{S \mathbf{Q_{13}} \downarrow} {n'}_{S \mathbf{k} \uparrow}
\big).
\label{eqn:ie}
\end{align}
Here we note that the variational parameters must satisfy
$
\alpha_{nm\ell}^{\mathbf{k_1}\mathbf{k_2}} (\mathbf{q}) 
= - \alpha_{mn\ell}^{\mathbf{k_2}\mathbf{k_1}} (\mathbf{q})
$
because $| \Psi_\mathbf{q} \rangle$ must be anti-symmetric under the exchange of 
$\uparrow$ particles. In addition, by introducing a new parameter set
$
\gamma_{n S}^\mathbf{q}(\mathbf{k}) = \sum_{m\ell \mathbf{k'}} 
\alpha_{nm\ell}^{\mathbf{k} \mathbf{k'}} (\mathbf{q}) 
m_{S \mathbf{k'} \uparrow} 
\ell_{S,\mathbf{q}-\mathbf{k}-\mathbf{k'}, \downarrow},
$
we bring Eq.~(\ref{eqn:ie}) to its somewhat familiar form
\begin{align}
\gamma_{n S}^\mathbf{q}(\mathbf{k}) = \frac{U}{N_c} & \sum_{m\ell S' \mathbf{k'}}
\frac{\ell_{S',\mathbf{q}-\mathbf{k}-\mathbf{k'}, \downarrow}^* 
\ell_{S,\mathbf{q}-\mathbf{k}-\mathbf{k'}, \downarrow} m_{S \mathbf{k'} \uparrow}}
{\varepsilon_{n \mathbf{k} \uparrow} + \varepsilon_{m \mathbf{k'} \uparrow} + 
\varepsilon_{\ell, \mathbf{q}-\mathbf{k}-\mathbf{k'}, \downarrow} - E_{3b}^\mathbf{q}}
\nonumber \\
& \times
\big[
m_{S' \mathbf{k'} \uparrow}^* \gamma_{n S'}^\mathbf{q}(\mathbf{k})
- n_{S' \mathbf{k} \uparrow}^* \gamma_{m S'}^\mathbf{q}(\mathbf{k'})
\big].
\label{eqn:gamma}
\end{align}
This is the multiband generalization of the three-body problem: it requires the solution 
of $N_b^2$ coupled integral equations for $\gamma_{n S}^\mathbf{q}(\mathbf{k})$. 
The well-known one-band result is recovered by setting the Bloch factors to unity and 
dropping the band as well as sublattice indices, i.e., it requires the solution of a single 
integral equation for $\gamma^\mathbf{q}(\mathbf{k})$~\cite{mattis86, orso10, orso11}.

In comparison the two-body bound states are determined by a set of self-consistency 
relations~\cite{iskin21, iskin22, orso21}
\begin{align}
\label{eqn:beta}
\beta_{S}^\mathbf{q} = \frac{U}{N_c} & \sum_{m\ell S' \mathbf{k}}
\frac{m_{S' \mathbf{k} \uparrow}^* \ell_{S',\mathbf{q}-\mathbf{k}, \downarrow}^* 
\ell_{S,\mathbf{q}-\mathbf{k}, \downarrow} m_{S \mathbf{k} \uparrow}}
{\varepsilon_{m \mathbf{k} \uparrow} + 
\varepsilon_{\ell, \mathbf{q}-\mathbf{k}, \downarrow} - E_{2b}^\mathbf{q}}
\beta_{S'}^\mathbf{q}
\end{align}
for a given total momentum $\mathbf{q}$ of the two particles. 
Note that Eq.~(\ref{eqn:beta}) is disguised in the first term of the second line in 
Eq.~(\ref{eqn:gamma}), and can be revealed by setting $\mathbf{k } = \mathbf{0}$ and 
$\varepsilon_{n \mathbf{k} \uparrow} = 0$ there. 
It is relatively much easier to solve Eq.~(\ref{eqn:beta}) by representing it as 
an $N_b \times N_b$ matrix for the $\beta_S^\mathbf{q}$ parameters, 
leading to $N_b$ bound-state solutions for a given $\mathbf{q}$.

\section{Numerical Implementation}
\label{sec:num}

Even though Eq.~(\ref{eqn:gamma}) is in the form of a set of coupled integral equations, 
we are interested only in $E_{3b}^\mathbf{q}$ as a function of $\mathbf{q}$ but not 
the variational parameters $\alpha_{nm\ell}^{\mathbf{k} \mathbf{k'}}(\mathbf{q})$ or 
$\gamma_{n S}^\mathbf{q}(\mathbf{k})$. 
For this reason it is possible to extract $E_{3b}^\mathbf{q}$ from Eq.~(\ref{eqn:gamma}) 
without the need of its explicit solutions.
Here we describe our numerical recipe for those lattices with a two-point basis, i.e., a two-band 
lattice with $N_b = 2$. Its generalization to arbitrary $N_b$ is obvious.

First we note that Eq.~(\ref{eqn:gamma}) has the generic form,
$
\gamma_{n S}^\mathbf{q}(\mathbf{k}) = 
\sum_{S'} f_{nS; nS'}^{\mathbf{q} \mathbf{k}} \gamma_{n S'}^\mathbf{q}(\mathbf{k}) 
+ \sum_{mS' \mathbf{k'}} g_{nS; mS'}^{\mathbf{q} \mathbf{k} \mathbf{k'}} \gamma_{m S'}^\mathbf{q}
(\mathbf{k'}),
$
and its coefficients $f_{nS; nS'}^{\mathbf{q} \mathbf{k}}$ and 
$g_{nS; mS'}^{\mathbf{q} \mathbf{k} \mathbf{k'}}$ are stored as
\begin{align}
f_{nS; nS'}^{\mathbf{q} \mathbf{k}} &= \frac{U}{N_c} \sum_{m\ell \mathbf{k'}}
\frac{\ell_{S',\mathbf{q}-\mathbf{k}-\mathbf{k'}, \downarrow}^* 
\ell_{S,\mathbf{q}-\mathbf{k}-\mathbf{k'}, \downarrow} 
m_{S \mathbf{k'} \uparrow} m_{S' \mathbf{k'} \uparrow}^*}
{\varepsilon_{n \mathbf{k} \uparrow} + \varepsilon_{m \mathbf{k'} \uparrow} + 
\varepsilon_{\ell, \mathbf{q}-\mathbf{k}-\mathbf{k'}, \downarrow} - E_{3b}^\mathbf{q}},
\\
g_{nS; mS'}^{\mathbf{q} \mathbf{k} \mathbf{k'}} &= -\frac{U}{N_c} \sum_{\ell}
\frac{\ell_{S',\mathbf{q}-\mathbf{k}-\mathbf{k'}, \downarrow}^* 
\ell_{S,\mathbf{q}-\mathbf{k}-\mathbf{k'}, \downarrow} 
m_{S \mathbf{k'} \uparrow} n_{S' \mathbf{k} \uparrow}^*}
{\varepsilon_{n \mathbf{k} \uparrow} + \varepsilon_{m \mathbf{k'} \uparrow} + 
\varepsilon_{\ell, \mathbf{q}-\mathbf{k}-\mathbf{k'}, \downarrow} - E_{3b}^\mathbf{q}}.
\end{align}
Then we define an $N_b^2$-component vector
$
\boldsymbol{\gamma}^\mathbf{q} (\mathbf{k}) = 
\begin{bmatrix}
\gamma_{1A}^\mathbf{q}(\mathbf{k}) & \gamma_{1B}^\mathbf{q}(\mathbf{k}) &
\gamma_{2A}^\mathbf{q}(\mathbf{k}) & \gamma_{2B}^\mathbf{q}(\mathbf{k})
\end{bmatrix}^\mathrm{T}
$
for a given $\mathbf{q}$ and $\mathbf{k}$, where $n = \{1,2\}$ refers to the band indices,
$S = \{A,B\}$ refers to the sublattices and $\mathrm{T}$ is the transpose, 
and recast Eq.~(\ref{eqn:gamma}) as 
$
\boldsymbol{\gamma}^\mathbf{q} (\mathbf{k}) 
= F^{\mathbf{q} \mathbf{k}} \boldsymbol{\gamma}^\mathbf{q} (\mathbf{k}) 
+ \sum_\mathbf{k'} G^{\mathbf{q} \mathbf{k} \mathbf{k'}} 
\boldsymbol{\gamma}^\mathbf{q} (\mathbf{k'}).
$
Here $F^{\mathbf{q} \mathbf{k}}$ and $G^{\mathbf{q} \mathbf{k} \mathbf{k'}}$ 
are $N_b^2 \times N_b^2$ matrices, e.g., 
\begin{align}
F^{\mathbf{q} \mathbf{k}} &= 
\begin{pmatrix}
f_{1A;1A}^{\mathbf{q} \mathbf{k}} &
f_{1A;1B}^{\mathbf{q} \mathbf{k}} &
0 & 0 \\
f_{1B;1A}^{\mathbf{q} \mathbf{k}} &
f_{1B;1B}^{\mathbf{q} \mathbf{k}} &
0 & 0 \\
0 & 0 &
f_{2A;2A}^{\mathbf{q} \mathbf{k}} &
f_{2A;2B}^{\mathbf{q} \mathbf{k}} \\
0 & 0 &
f_{2B;2A}^{\mathbf{q} \mathbf{k}} &
f_{2B;2B}^{\mathbf{q} \mathbf{k}}
\end{pmatrix},
\\
G^{\mathbf{q} \mathbf{k} \mathbf{k'}} &= 
\begin{pmatrix}
g_{1A;1A}^{\mathbf{q} \mathbf{k} \mathbf{k'}} &
g_{1A;1B}^{\mathbf{q} \mathbf{k} \mathbf{k'}} &
g_{1A;2A}^{\mathbf{q} \mathbf{k} \mathbf{k'}} &
g_{1A;2B}^{\mathbf{q} \mathbf{k} \mathbf{k'}} \\
g_{1B;1A}^{\mathbf{q} \mathbf{k} \mathbf{k'}} &
g_{1B;1B}^{\mathbf{q} \mathbf{k} \mathbf{k'}} &
g_{1B;2A}^{\mathbf{q} \mathbf{k} \mathbf{k'}} &
g_{1B;2B}^{\mathbf{q} \mathbf{k} \mathbf{k'}} \\
g_{2A;1A}^{\mathbf{q} \mathbf{k} \mathbf{k'}} &
g_{2A;1B}^{\mathbf{q} \mathbf{k} \mathbf{k'}} &
g_{2A;2A}^{\mathbf{q} \mathbf{k} \mathbf{k'}} &
g_{2A;2B}^{\mathbf{q} \mathbf{k} \mathbf{k'}} \\
g_{2B;1A}^{\mathbf{q} \mathbf{k} \mathbf{k'}} &
g_{2B;1B}^{\mathbf{q} \mathbf{k} \mathbf{k'}} &
g_{2B;2A}^{\mathbf{q} \mathbf{k} \mathbf{k'}} &
g_{2B;2B}^{\mathbf{q} \mathbf{k} \mathbf{k'}}
\end{pmatrix},
\end{align}
when $N_b = 2$.
Finally we define an $N_c N_b^2$-component vector
$
\boldsymbol{\gamma}^\mathbf{q} = 
\begin{bmatrix}
\boldsymbol{\gamma}^\mathbf{q} (\mathbf{k_1}) & \boldsymbol{\gamma}^\mathbf{q} (\mathbf{k_2}) &
\dots & \boldsymbol{\gamma}^\mathbf{q} (\mathbf{k_{N_c}})
\end{bmatrix}^\mathrm{T}
$
for a given $\mathbf{q}$, where 
$
\mathbf{k} = \{ \mathbf{k_1}, \mathbf{k_2}, \dots, \mathbf{k_{N_c}} \},
$
corresponds to the mesh points in the first BZ, and recast Eq.~(\ref{eqn:gamma}) as
\begin{align}
\label{eqn:GF}
(\mathbf{G}^\mathbf{q} + \mathbf{F}^\mathbf{q}) \boldsymbol{\gamma}^\mathbf{q} = 
\boldsymbol{\gamma}^\mathbf{q}.
\end{align}
Here $\mathbf{G}^\mathbf{q}$ and $\mathbf{F}^\mathbf{q}$ are $N_c N_b^2 \times N_c N_b^2$ 
matrices, and they are formed, respectively, from $G^{\mathbf{q} \mathbf{k} \mathbf{k'}}$ 
and $F^{\mathbf{q} \mathbf{k}}$ matrices, i.e., 
\begin{align}
\mathbf{F}^\mathbf{q} &= \left(
\begin{array}{cccc}
   F^{\mathbf{q} \mathbf{k_1}} & 0 & \cdots & 0 
\\
   0 & F^{\mathbf{q} \mathbf{k_2}}  & \cdots & 0 
\\
   \vdots & \vdots & \ddots & \vdots   
\\
 0 & 0 & \cdots & F^{\mathbf{q} \mathbf{k_{N_c}}}   
\end{array} \right),
\\
\mathbf{G}^\mathbf{q} &= \left(
\begin{array}{cccc}
   G^{\mathbf{q} \mathbf{k_1} \mathbf{k_1}} & G^{\mathbf{q} \mathbf{k_1} \mathbf{k_2}} & \cdots & G^{\mathbf{q} \mathbf{k_1} \mathbf{k_{N_c}}} 
\\
   G^{\mathbf{q} \mathbf{k_2} \mathbf{k_1}} & G^{\mathbf{q} \mathbf{k_2} \mathbf{k_2}} & \cdots & G^{\mathbf{q} \mathbf{k_2} \mathbf{k_{N_c}}} 
\\
   \vdots & \vdots & \ddots & \vdots   
\\
   G^{\mathbf{q} \mathbf{k_{N_c}} \mathbf{k_1}} & G^{\mathbf{q} \mathbf{k_{N_c}} \mathbf{k_2}} & \cdots & G^{\mathbf{q} \mathbf{k_{N_c}} \mathbf{k_{N_c}}} 
\end{array} \right).
\end{align}
Note that both matrices are Hermitian because
$
f_{nS;nS'}^{\mathbf{q} \mathbf{k}} = (f_{nS';nS}^{\mathbf{q} \mathbf{k}})^*
$
and
$
g_{nS;mS'}^{\mathbf{q} \mathbf{k} \mathbf{k'}} = (g_{mS';nS}^{\mathbf{q} \mathbf{k'} \mathbf{k}})^*.
$

Thus the three-body problem reduces to the solutions of an eigenvalue problem defined 
by Eq.~(\ref{eqn:GF}). It can be solved numerically by iterating $E_{3b}^\mathbf{q}$ until one 
of the eigenvalues of $\mathbf{G}^\mathbf{q} + \mathbf{F}^\mathbf{q}$ becomes exactly 1.
Here we use a hybrid root-finding algorithm which combines the bisection and secant 
methods. Depending on the initial choice of $E_{3b}^\mathbf{q}$, the iterative approach 
may converge to one of the higher-energy bound state or scattering-state solutions. 
In this paper we are interested in the lowest bound state with minimum allowed 
$E_{3b}^\mathbf{q}$ for a given $\mathbf{q}$. Thus, by choosing a lower and lower 
initial $E_{3b}^\mathbf{q}$ value, we made sure that there does not exist a solution 
with lower energy.

Having discussed the theoretical analysis of the three-body problem in a generic 
multiband lattice, next we apply our numerical recipe to the sawtooth lattice.

\section{Sawtooth Lattice}
\label{sec:saw}

\begin{figure}[htbp]
\includegraphics[scale=0.29]{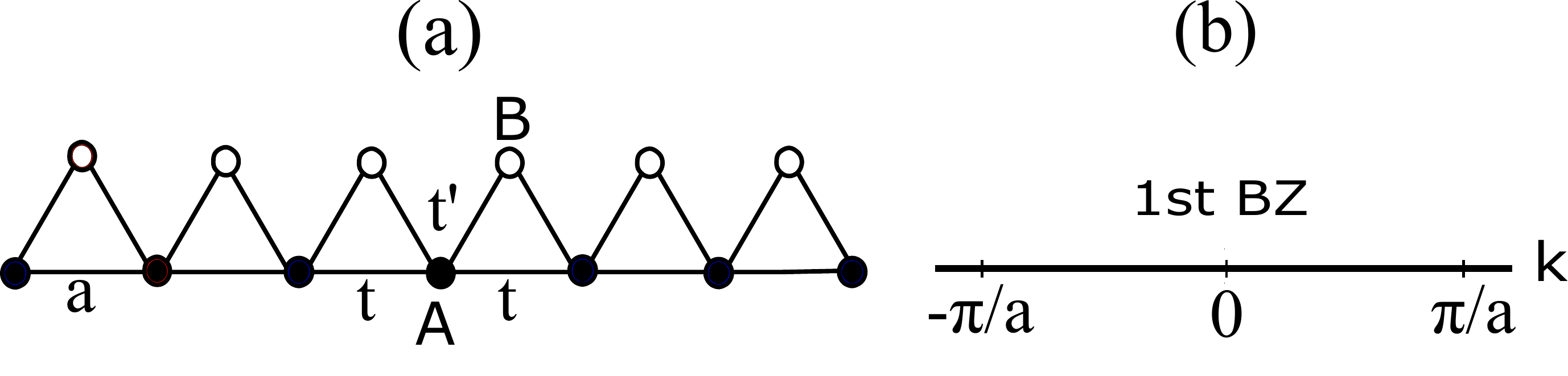}
\vskip 0.5cm
\centerline{\scalebox{0.31}{\includegraphics{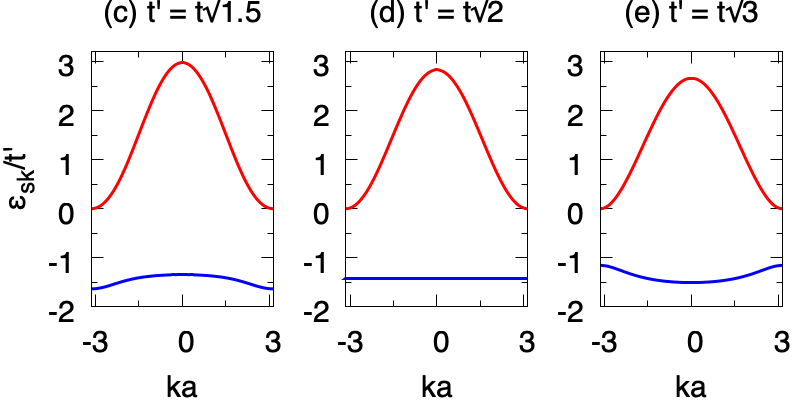}}}
\caption{\label{fig:bs}
Sawtooth lattice is a linear-chain model with a two-point basis (a), and its first Brillouin zone 
lies on a straight line (b). Typical band structures (c-e) feature a flat band with energy
$\varepsilon_{-, k} = -2t$ when $t' = t\sqrt{2}$.
}
\end{figure}

In part due to its flat band and one-dimensional simplicity, the sawtooth lattice (also called
the one-dimensional Tasaki lattice) is one of the well-studied lattice models in 
recent literature~\cite{zhang15, pyykonen21, chan21, orso21}. 
It is a linear chain of equidistant lattice points (with spacing $a$) that are attached 
with a two-point basis ($A$ and $B$ sites) as shown in Fig.~\ref{fig:bs}(a), 
and its first BZ lies between $-\pi/a$ and $\pi/a$ as shown in Fig.~\ref{fig:bs}(b).

In this paper we allow hopping processes between nearest-neighbor sites only, and set 
$
t_{Aj;Ai}^\sigma = - t
$
with $j = i \pm 1$ and $t \ge 0$,
$
t_{Bj;Bi}^\sigma = 0
$
and 
$
t_{Bi;Ai}^\sigma = t_{Bj;Ai}^\sigma = - t'
$
with $j = i-1$ and $t' \ge 0$. It is called the zigzag model when $t_{Bj;Bi}^\sigma \ne t''$ 
for $j = i \pm 1$. Then the single-particle Hamiltonian can be written as
\begin{align}
H_\sigma = \sum_k
\begin{pmatrix}
c_{A k \sigma}^\dagger & c_{B k \sigma}^\dagger
\end{pmatrix}
\begin{pmatrix}
d_k^0 + d_k^z & d_k^x - id_k^y  \\ 
d_k^x + id_k^y & d_k^0 - d_k^z  \\ 
\end{pmatrix}
\begin{pmatrix}
c_{A k \sigma} \\ c_{B k \sigma}
\end{pmatrix},
\end{align}
where the wave vector $k \in \textrm{BZ}$, and the matrix elements are
$
d_k^0 = d_k^z = t \cos(k a),
$
$
d_k^x = t' + t' \cos(k a)
$
and
$
d_k^y = t' \sin(k a).
$
Thus the single-particle energy bands disperse as
$
\varepsilon_{s k \sigma} =  d_k^0 + s d_k
$
where $s = \pm$ labels the upper and lower bands, respectively, and 
$
d_k = \sqrt{(d_k^x)^2 + (d_k^y)^2 + (d_k^z)^2}.
$
The corresponding eigenvectors are determined by
$
s_{A k \sigma} = \langle A | s k \sigma \rangle = 
\frac{- d_k^x + id_k^y} {\sqrt{2d_k(d_k - sd_k^z)}}
$
and
$
s_{B k \sigma} = \langle B | s k \sigma \rangle = 
\frac{d_k^z - sd_k} {\sqrt{2d_k(d_k - sd_k^z)}}.
$
We illustrate typical band structures $\varepsilon_{s k} = \varepsilon_{s k \sigma}$ in 
Figs.~\ref{fig:bs}(c), \ref{fig:bs}(d) and~\ref{fig:bs}(e).
It is shown that while the lower band is flat with energy $\varepsilon_{-, k} = -2t$ 
when $t'/t = \sqrt{2}$, it has a positive (negative) curvature when $t'/t$ is greater 
(lesser) than $\sqrt{2}$.

\begin{figure}[htbp]
\centerline{\scalebox{0.31}{\includegraphics{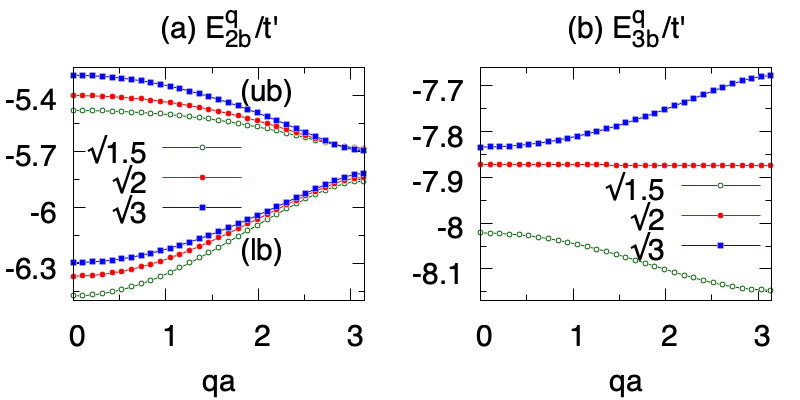}}}
\vskip 0.2cm
\centerline{\scalebox{0.31}{\includegraphics{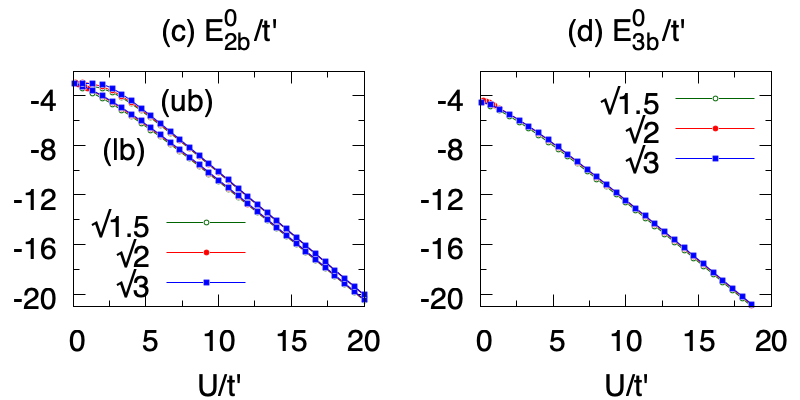}}}
\caption{\label{fig:23}
(a) There are two distinct two-body bound states $E_{2b}^q$ for a given total momentum 
$q$ of the two particles: upper branch (ub) and lower one (lb). 
The lower branch plays an important role in the stability of the trimers [see the discussion
around Eq.~(\ref{eqn:Etrim})].
(b) The energy of the lowest three-body bound state $E_{3b}^q$ as a function of total 
momentum $q$ of the three particles. 
While $E_{3b}^q$ of the flat-band case has a small dispersion that is similar in shape to that of 
the $\sqrt{1.5}$ case, it appears quite flat in the shown scale.
(c-d) $E_{2b}^0$ and $E_{3b}^0$ as a function of interaction. 
In all figures green-hollow-circles, red-filled-circles and blue-filled-squares 
correspond, respectively, to $t'/t = \{ \sqrt{1.5}, \sqrt{2}, \sqrt{3} \}$ that
are illustrated in Figs.~\ref{fig:bs}(c), \ref{fig:bs}(d) and~\ref{fig:bs}(e).
They are on top of each other in (c) and (d) except for the weak-coupling limit.
In addition we set $U = 5t'$ in (a) and (b).
}
\end{figure}

In Figs.~\ref{fig:23}(a) and~\ref{fig:23}(b), we set $U = 5t'$, and present, respectively, the 
corresponding solutions for the two-body ($E_{2b}^q$) and the three-body ($E_{3b}^q$)
bound states as a function of $q$. Here $q$ stands, respectively, for the total momentum 
of two and three particles involved. Since $N_b = 2$, there are two distinct $E_{2b}^q$ 
solutions for a given $q$: upper branch (ub) and lower one (lb). The lower branch
plays an important role in the stability of the trimers as discussed below.
We find that the $q$-dependences of $E_{2b}^q$ are qualitatively similar 
to each other for all three hoppings 
considered in Figs.~\ref{fig:bs}(c), \ref{fig:bs}(d) and~\ref{fig:bs}(e). In contrast, the 
$q$-dependences of $E_{3b}^q$ are quite distinct: while it has a positive (negative) curvature
near the origin (edge) of the BZ when $t'/t = \sqrt{3}$, it has a negative (positive) curvature
near the origin (edge) of the BZ when $t'/t = \sqrt{1.5}$. We also find that $E_{3b}^q$ of the 
flat-band case has a small dispersion that is similar in shape to that of the $\sqrt{1.5}$ case, 
but it appears quite flat in the shown scale. Its bandwidth $\sim 0.001204t'$ starts from 
$-7.872373 t'$ at $q = 0$ and decreases to $-7.873577 t'$ at $q = \pi/a$. 
In the low-$qa$ limit we find the following fitting functions for Fig.~\ref{fig:23}(b): 
$
E_{3b}^q/t' \approx - 8.021633  - 0.0235 a^2 q^2 
$
in the range $qa \lesssim 1$ when $t'/t = \sqrt{1.5}$,
$
E_{3b}^q/t' \approx - 7.872373 - 0.000635 a^2 q^2  
$
in the range $qa \lesssim 0.5$ when $t'/t = \sqrt{2}$, and
$
E_{3b}^q/t' \approx - 7.833807 + 0.0221 a^2 q^2  
$
in the range $qa \lesssim 1$ when $t'/t = \sqrt{3}$. All of these results are obtained 
with $N_c = 100$ mesh points in the BZ, and we checked that increasing it to 
$N_c = 200$ makes minor changes. 
Thus the flatness of the $E_{3b}^q$ when $t'/t = \sqrt{2}$ is partly caused by the large 
effective-mass of the three-body bound states.
In Figs.~\ref{fig:23}(c) and~\ref{fig:23}(d), we set $q = 0$, and present, respectively, 
$E_{2b}^0$ and $E_{3b}^0$ as a function of $U$. 
They appear on top of each other for different values of $t$ except for the weak-coupling limit.

\begin{figure}[htbp]
\centerline{\scalebox{0.3}{\includegraphics{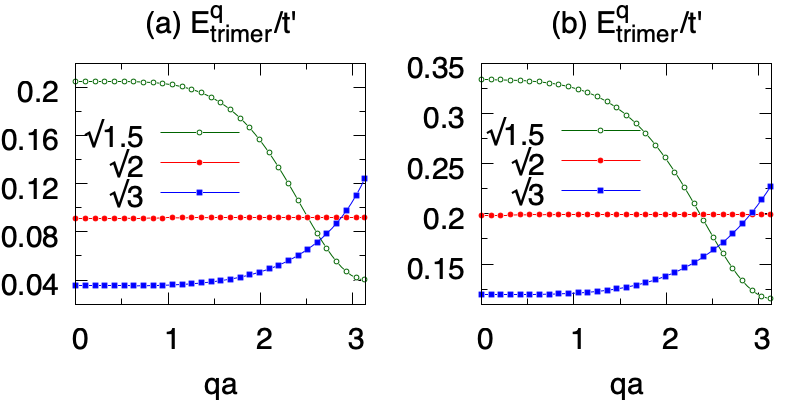}}}
\vskip 0.2cm
\centerline{\scalebox{0.3}{\includegraphics{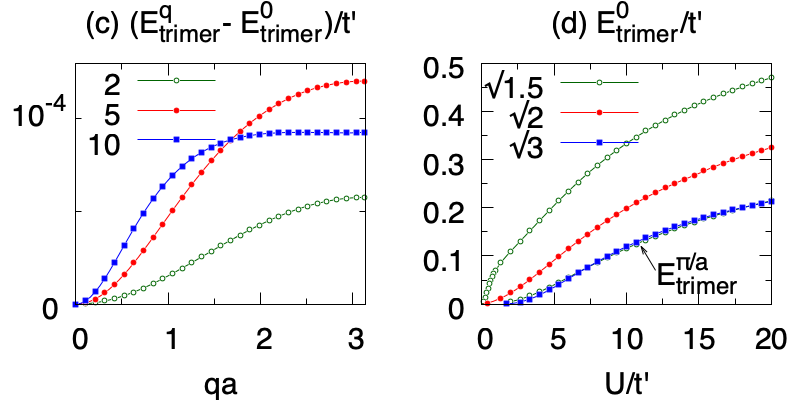}}}
\caption{\label{fig:trim}
(a - b) Binding energy of the three-body bound state $E_\textrm{trimer}^q$ as a function of 
total momentum $q$ of the three particles when $U = 5t'$ and $U = 10t'$, respectively. 
While $E_\textrm{trimer}^q$ of the flat-band case has a small dispersion 
with a positive (upward) curvature near the origin, it appears quite flat in the shown scale. 
This is illustrated in (c) where $E_\textrm{trimer}^q - E_\textrm{trimer}^0$ is shown as a 
function of $q$ for $U/t' = \{2, 5, 10\}$ when $t' = t\sqrt{2}$.
(d) $E_\textrm{trimer}^0$ as a function of interaction.
$E_\textrm{trimer}^{\pi/a}$ is also shown for $t' = t\sqrt{1.5}$ but it is barely visible
since it overlaps with the $E_\textrm{trimer}^0$ of $t' = t\sqrt{3}$ in most parts.
Note that $E_\textrm{trimer}^0$ of the flat-band case is consistent with the very 
recent DMRG results (see their Fig.~10)~\cite{orso21}.
In (a), (b) and (d) the green-hollow-circles, red-filled-circles and blue-filled-squares 
correspond, respectively, to $t'/t = \{ \sqrt{1.5}, \sqrt{2}, \sqrt{3} \}$ that
are illustrated in Figs.~\ref{fig:bs} and~\ref{fig:23}. 
}
\end{figure}

In order to be observed, a three-body bound state (trimer) must be energetically stable 
against two distinct dissociation mechanisms~\cite{shi14}: 
(i) free-atom dissociation threshold where the trimer decays into two free spin-$\uparrow$ 
particles and a free spin-$\downarrow$ particle,
and (ii) atom-dimer dissociation threshold where the trimer decays into a two-body bound 
state (dimer) and a free spin-$\uparrow$ particle.
Since the former mechanism requires higher-energy processes in the parameter regime of 
interest in our numerical calculations, it is the second mechanism that determines the 
binding energy $E_\textrm{trimer}^\mathbf{q}$ of the trimers.
For this reason we define $E_\textrm{trimer}^\mathbf{q}$ with respect to the 
atom-dimer dissociation threshold as
\begin{align}
\label{eqn:Etrim}
E_\textrm{trimer}^\mathbf{q} = -E_{3b}^\mathbf{q} 
+ \min \{ E_{2b}^\mathbf{q'} + \varepsilon_{n, \mathbf{q} - \mathbf{q'}, \uparrow} \}.
\end{align}
In Fig.~\ref{fig:trim}(a) we set $U = 5t'$, and present the resultant $E_\textrm{trimer}^q$ 
as a function of $q$ for the corresponding data shown in Figs.~\ref{fig:23}(a) and~\ref{fig:23}(b).
We found very similar results for most of the parameter regimes of interest here, 
e.g., $U = 10t'$ is shown in Fig.~\ref{fig:trim}(b).
In particular, in the flat-band case when $t'/t = \sqrt{2}$, the atom-dimer dissociation 
threshold is given by
$
\min \{ E_{2b}^{q'} + \varepsilon_{n, q - q', \uparrow} \}
= \min \{ E_{2b}^{q'} \} - 2t
= E_{2b}^0 - 2t.
$
This is because the dimer ground-state is at $q = 0$, and $E_{2b}^0$ is the 
minimum of the lower branch in the two-body problem.
Thus while $E_\textrm{trimer}^q$ of the flatband case has a small dispersion with a 
positive (upward) curvature coming from $-E_{3b}^\mathbf{q}$, it appears quite flat
in the shown scale. To illustrate its dispersive nature we present 
$
E_\textrm{trimer}^q - E_\textrm{trimer}^0
$ 
in Fig.~\ref{fig:trim}(c) for 
$
U/t' = \{ 2, 5, 10\},
$
where 
$
E_\textrm{trimer}^0/t' \sim \{ 0.023, 0.091, 0.20 \},
$
respectively. This figure suggests that $E_\textrm{trimer}^q$ may have a sizeable 
dispersion only in the weak-coupling limit when $E_\textrm{trimer}^0$ is small. 
Unfortunately our numerical accuracy becomes unreliable in this limit, and we 
could not fully resolve this point. This is because as the size of the trimers 
(in real space) is expected to increase dramatically in the 
$
E_\textrm{trimer}^0/t' \to 0
$ 
limit, their precise calculation requires a much larger lattice size, i.e., 
one must choose larger and larger number of unit cells $N_c \to \infty$ 
as $U/t' \to 0$.

Furthermore Figs.~\ref{fig:trim}(a) and~\ref{fig:trim}(b) shows that while the binding-energy
of the ground-state trimer is at $q = \pi/a$ when $t'/t = \sqrt{1.5}$, it is at $q = 0$ when 
$t'/t = \sqrt{2}$ or $\sqrt{3}$. The origin of this difference can be traced back 
to the location of the single-particle ground state, i.e., see the corresponding band 
structures in Figs.~\ref{fig:bs}(c), \ref{fig:bs}(d) and~\ref{fig:bs}(e), respectively.
In order to reveal the fate of trimer states as a function of $U$, we set $q = 0$, 
and present the resultant $E_\textrm{trimer}^0$ in Fig.~\ref{fig:trim}(d) for the 
corresponding data shown in Figs.~\ref{fig:23}(c) and~\ref{fig:23}(d).
We also show $E_\textrm{trimer}^{\pi/a}$ for the $t' = t\sqrt{1.5}$ case but it is 
barely visible since it overlaps with the $E_\textrm{trimer}^0$ of $t' = t\sqrt{3}$ in 
most parts. In addition $E_\textrm{trimer}^0$ of the $t' = t\sqrt{1.5}$ case is shown 
for completeness.
First of all it is delightful to note that $E_\textrm{trimer}^0$ of the flat-band 
case seems to be in very good agreement with the recent DMRG results, 
i.e., compare it with Fig.~10 of~\cite{orso21}.
In this case our numerical findings suggest that there exist trimer states that are 
energetically stable for all interaction strengths including the weak-coupling limit 
no matter how small $U/t'$ is.

On the other hand, when $t'/t$ deviates from $\sqrt{2}$, there seems to be a finite 
threshold in the $U/t' \to 0$ limit. For instance $E_\textrm{trimer}^0$ of the 
$t'/t = \sqrt{3}$ case is shown in Fig.~\ref{fig:trim}(d), and we also verified it to be 
the case for the $t'/t = \sqrt{6}$ case but it is not presented.
In addition $E_\textrm{trimer}^{\pi/a}$ of the $t'/t = \sqrt{1.5}$ case is again shown 
in Fig.~\ref{fig:trim}(d), and we also verified it to be the case for the $t'/t = 1$ case 
but is again not presented.
It is numerically challenging to pinpoint the exact location of the interaction thresholds 
in the $U/t' \to 0$ limit since the binding energy of the ground-state trimers, i.e., 
$E_\textrm{trimer}^0$ or $E_\textrm{trimer}^{\pi/a}$, gradually approaches to zero 
with a long tail. However we observe that the thresholds tend to increase further and 
further as a function of increasing deviation from the flat-band limit $t'/t = \sqrt{2}$, i.e.,
the threshold for the $t'/t = 1$ case is considerably higher than that of $t'/t = \sqrt{1.5}$
and the threshold for the $t'/t = \sqrt{6}$ case is considerably higher than that of $t'/t = \sqrt{3}$.
Our naive expectation is that the sawtooth model must recover the linear-chain model 
in either (i) the $t'/t \gg 1$ or (ii) the $t'/t \ll 1$ limit. In fact, in agreement with our 
numerical results, stable trimers are known not to be allowed in a single-band 
linear-chain model~\cite{mattis86, orso10, orso11}. 
Thus our results establish that the formation of stable trimers is a genuine multiband 
effect mediated by the interband transitions.

\section{Conclusion}
\label{sec:conc}

To summarize here we solved the three-body problem in a generic multiband Hubbard
model, and reduced it to an eigenvalue problem for the dispersion of the trimer states.
As an illustration we applied our theory to the sawtooth lattice with a two-point basis,
and showed that the trimer states are allowed in a broad range of model parameters.
This finding is in sharp contrast with the single-band linear-chain 
model~\cite{mattis86, orso10, orso11} and it is in very good agreement with 
the recent DMRG results~\cite{orso21}. 
In addition we found that the trimers have a nearly-flat dispersion with a negligible 
bandwidth when formed in a flat band, which is unlike the highly-dispersive spectrum 
of its dimers. As an outlook our generic results may find direct applications in 
higher-dimensional lattices with more complicated lattice geometries and band 
structures~\cite{mizoguchi19}. For instance the fate of trimers in a Kagome lattice
could be an interesting problem~\cite{iskin22}. Such an analysis would reveal not only 
the impact of higher bands on the trimer states but also the role played by the lattice 
dimensionality.
Furthermore it is a straightforward task to extend our approach and analyze the 
nature of trimer states with three identical bosons in the presence of multiple Bloch
bands~\cite{mattis86, valiente10}. 

As a final remark we have recently generalized our approach to the ($N+1$)-body 
problem in a generic multiband lattice, and derived the integral equations for the 
bound states of $N$ spin-$\uparrow$ fermions and a spin-$\downarrow$ fermion 
due to an onsite attraction in between~\cite{iskin22}. Our numerical calculations 
for the $N = 3$ case shows that the tetramer states are also allowed in the 
sawtooth lattice, e.g., they also have a nearly-flat dispersion with a negligible 
bandwidth when formed in a flat band. It turns out larger cluster states, 
i.e., pentamers and beyond, are also possible in this system but, unfortunately, 
one may have to resort to a high-performance computer to solve the resultant 
matrices when $N \ge 4$. They are numerically very expensive and well beyond our 
current capacity.

\begin{acknowledgments}
The author acknowledges funding from T{\"U}B{\.I}TAK.
\end{acknowledgments}

\end{document}